\documentclass[aps, prd, twocolumn, superscriptaddress,preprintnumbers,nofootinbib]{revtex4}

\usepackage{bm,graphicx}

\usepackage{dcolumn}

\usepackage{amssymb}

\begin{document}

\newcommand{\fd}{f_{10}}

\newcommand{\As}{A_{\mathrm{s}}}

\newcommand{\At}{A_{\mathrm{t}}}

\newcommand{\ns}{n_{\mathrm{s}}}

\newcommand{\nt}{n_{\mathrm{t}}}

\newcommand{\Obhh}{\Omega_{\mathrm{b}}h^{2}}

\newcommand{\Omhh}{\Omega_{\mathrm{m}}h^{2}}

\newcommand{\Ol}{\Omega_{\Lambda}}

\newcommand{\half}{\frac{1}{2}}

\newcommand{\CMB}{\textsc{CMB }}

\newcommand{\CMBEASY}{\textsc{CMBeasy }}

\newcommand{\TT}{\textsc{tt }}

\newcommand{\TE}{\textsc{te }}

\newcommand{\EE}{\textsc{ee }}

\newcommand{\BB}{\textsc{bb }}

\newcommand{\WMAP}{\textsc{WMAP }}

\newcommand{\UETC}{\textsc{uetc }}

\newcommand{\UETCs}{\textsc{uetc}s}

\newcommand{\ETC}{\textsc{etc }}

\newcommand{\ETCs}{\textsc{etc}s }

\newcommand{\HK}{\textsc{HKP }}

\newcommand{\BBN}{\textsc{BBN }}

\newcommand{\mcmc}{\textsc{MCMC }}

\newcommand{\clover}{\textsc{C}$\ell$\textsc{over}}

\renewcommand{\d}{{\partial}}

\newcommand{\be}{\begin{equation}}

\newcommand{\ee}{\end{equation}}

\newcommand{\bea}{\begin{eqnarray}}

\newcommand{\eea}{\end{eqnarray}}

\newcommand{\mbh}[1]{\textbf{#1}}

\preprint{NSF-KITP-08-29}
\title{On the degeneracy between primordial tensor modes and cosmic strings in future CMB data from Planck} 

\newcommand{\addressSussex}{Department of Physics \&
Astronomy, University of Sussex, Brighton, BN1 9QH, United Kingdom}

\author{Jon Urrestilla}
\affiliation{\addressSussex}
\author{Pia Mukherjee}
\affiliation{\addressSussex}
\author{Andrew R.~Liddle}
\affiliation{\addressSussex}

\author{Neil Bevis}
\affiliation{Theoretical Physics, Blackett Laboratory, Imperial
  College, London, SW7 2BZ, United Kingdom}

\author{Mark Hindmarsh}
\affiliation{\addressSussex}

\author{Martin Kunz}
\affiliation{\addressSussex}

\date{\today}

\begin{abstract}
While observations indicate that the predominant source of cosmic
inhomogeneities are adiabatic perturbations, there are a variety of
candidates to provide auxiliary trace effects, including
inflation-generated primordial tensors and cosmic defects which both
produce B-mode cosmic microwave background (CMB) polarization. We
investigate whether future experiments may suffer confusion as to the
true origin of such effects, focusing on the ability of Planck to 
distinguish tensors from cosmic strings, and
show that there is no significant degeneracy.
\end{abstract}

\maketitle


\section{Introduction}

Cosmological probes are reaching a sensitivity where they are able to
meaningfully constrain models of the early Universe. Data compilations
including the Wilkinson Microwave Anisotropy Probe five-year (WMAP5) data
  \cite{Nolta:2008ih}
 already indicate that the dominant source of
inhomogeneities are primordial adiabatic scalar
perturbations \cite{komatsu}. However, there remains
 room for low-level contributions
from other sources, for instance isocurvature perturbations, and
the discovery of such trace effects may be essential to enhance the
limited information available via the adiabatic scalars. Of particular
interest are primordial tensor perturbations, believed to be generated
by inflation alongside the scalars, and also cosmic defects.

Cosmological data may even be able to constrain
string/M-theory, the current dominant unification paradigm.  
There have been attempts to try to get direct information about string
theory from cosmology. For example, it may be possible
 to infer the topology and geometry
of the Calabi--Yau space in which the extra dimension are compactified
\cite{Bean:2007eh}.  Without going into
the model-dependent assumptions, a fairly general prediction from
string cosmology seems to be that the level of primordial
gravitational waves, given by the tensor-to-scalar ratio $r$, is very
low ($r \ll 10^{-3}$, in some cases even $r\sim 10^{-23})$; as
emphasized by Kallosh et al.~\cite{Kallosh:2007wm} there is no known
inflationary model coming from string theory which predicts measurably
high primordial tensor modes. Thus, a future detection of $r$ in the
accessible range $r\gtrsim10^{-2}$--$10^{-3}$ would present a tough
challenge for string cosmology.

Another typical prediction of string cosmology is the production of
cosmic strings \cite{Majumdar:2002hy}. These strings
can be fundamental strings or D1 branes (or D-branes with $D-1$
dimensions wrapped in the extra dimensions) left over from brane
inflation.  Alternatively it can be argued that D-term strings are the
low-energy effective cousins of D-strings
\cite{Dvali:2003zh}. The dynamics of a system consisting of F- (fundamental)
and D-strings is an evolving field \cite{HS06}, and more study is needed to
have a consistent picture of such a network.

Strings produced after inflation \cite{string} will also generate 
Cosmic Microwave Background (CMB) anisotropies
\cite{Wyman:2005tu,bevis,urrestilla}, which can be parametrized by an
amplitude $G\mu$, where $G$ is the gravitational constant and $\mu$ is
the string tension.  This poses the question: in the event
of a future CMB experiment detecting some `extra' ingredient beyond a
primordial (scalar) inflationary spectrum, would its identification as
inflationary tensors be secure, or might cosmic strings have generated
a signal mistaken as primordial tensors? The interpretation of future
observations is clearly contingent on being able to make the right
model assumptions in fitting to data.  The aim of this paper is to
answer this simple question for the specific case of the Planck
satellite, due to be launched within the next year.

We remind the reader that both tensors and strings produce a
`primordial' B-mode polarization spectrum
\cite{Seljak:2006hi,Bevis:2007qz}, with fairly different
spectra. Unlike the other CMB spectra, these are not subdominant to
that from the primordial scalars, which is generated only indirectly
through lensing of E-modes into B-modes.  In principle, ground-based
and suborbital B-mode experiments would be more sensitive to both
tensors and strings. For example, a null detection by \clover\ would
give very tight constraints on the amount of strings possible
\cite{Bevis:2007qz}.  Nevertheless, the launch of Planck is imminent
and we will show that Planck alone is enough to answer our question in
a fairly definitive manner.

\section{Method}

In order to investigate the possible degeneracy between tensors and
cosmic strings, we created simulated Planck data for a few different
cosmologies. We include the temperature (TT) and E- and B-mode
 polarization spectra
(TE, EE, and BB) from three temperature channels with specification
similar to the HFI channels of frequency 100 GHz, 143 GHz, and 217
GHz, and one 143 GHz polarization channel, following the current
Planck documentation \cite{planck}.
We use a fiducial model close to the \WMAP best-fit flat $\Lambda$CDM
model, with $\Omega_{\rm b}h^2=0.022$, $\Omega_{\rm c}h^2=0.105$,
$H_0=73\ {\rm km} {\rm s}^{-1} {\rm Mpc}^{-1}$, $\tau=0.09$, $n_{\rm s}=0.96$, and $A_{\rm s}^{2} = 2.35\times
10^{-9}$.
The parameters $r$ and $G\mu$
take on various values.  The likelihood is
constructed assuming a fractional sky coverage of 0.8, and up to a maximum
multipole of 2000. We use CosmoMC \cite{Lewis:2002ah} to obtain
parameter confidence contours.

The CMB anisotropies created by cosmic strings were also included in
the simulated data. For this we use the results from
Refs.~\cite{bevis,Bevis:2007qz},\footnote{ Refs.~\cite{bevis,Bevis:2007qz}
 employed a code in which a  bug has been discovered, and this had a
 small effect in Ref.~\cite{Bevis:2007gh} since it used their results
 directly (see the respective errata). Here we have used the corrected power 
spectra from Refs.~\cite{bevis,Bevis:2007qz} and quote the corrected results
 from Ref.~\cite{Bevis:2007gh}.} for both temperature and
polarization.  These CMB anisotropies are obtained from a
field-theoretical approach to cosmic strings, simulating the Abelian
Higgs model on a lattice. The energy--momentum tensor corresponding
to the cosmic strings is extracted and its unequal
time correlators (\UETCs) computed 
\cite{Durrer:2001cg} and then a modified version of CMBeasy \cite{Doran:2003sy} yields the CMB power spectra. We follow Ref. \cite{Bevis:2004} and use this subdominant string contribution calculated at only a single cosmology, which gives a negligible degradation of the likelihood values we obtain.

In the end this string contribution, scaled by an amplitude $G\mu$, is
simply added to the other spectra. In turn, $G\mu$ can be related to
$\fd$, which measures the fractional contribution of strings to the
total TT power spectrum at multipole $\ell=10$. Previous work
\cite{Wyman:2005tu,Battye:2006pk} constraining 
the amount of cosmic strings allowed from current \CMB
data \cite{cmb}
suggests that not only is a fair amount of
string allowed, but actually about 10\% of strings is preferred 
\cite{Bevis:2007gh}
($\fd\sim0.1$, $G\mu\sim0.8\times 10^{-6}$
in the Abelian Higgs
model) by a
$\Delta\chi^2=-3.5$. Using Bayesian evidence for model comparison, a 
logarithmic
 evidence difference of $1.8\pm0.2$
is obtained between a model
with strings with fixed $n_{\rm s}=1$, and the concordance model. 
In this sense, we may say that strings are preferred to tilt by the CMB 
data.\footnote{Our calculations predated the release of the
5-year WMAP data.} 
Allowing $n_{\rm s}$ to
deviate from unity, including constraints from Big Bang
Nucleosynthesis \cite{Kirkman:2003uv} and the Hubble Key Project
\cite{Freedman:2000cf} all reduce the case for strings:
an  upper bound of $\fd<0.10$
on the fraction of power due to strings is obtained.

\begin{figure}[t]
\resizebox{0.9\columnwidth}{!}{\includegraphics{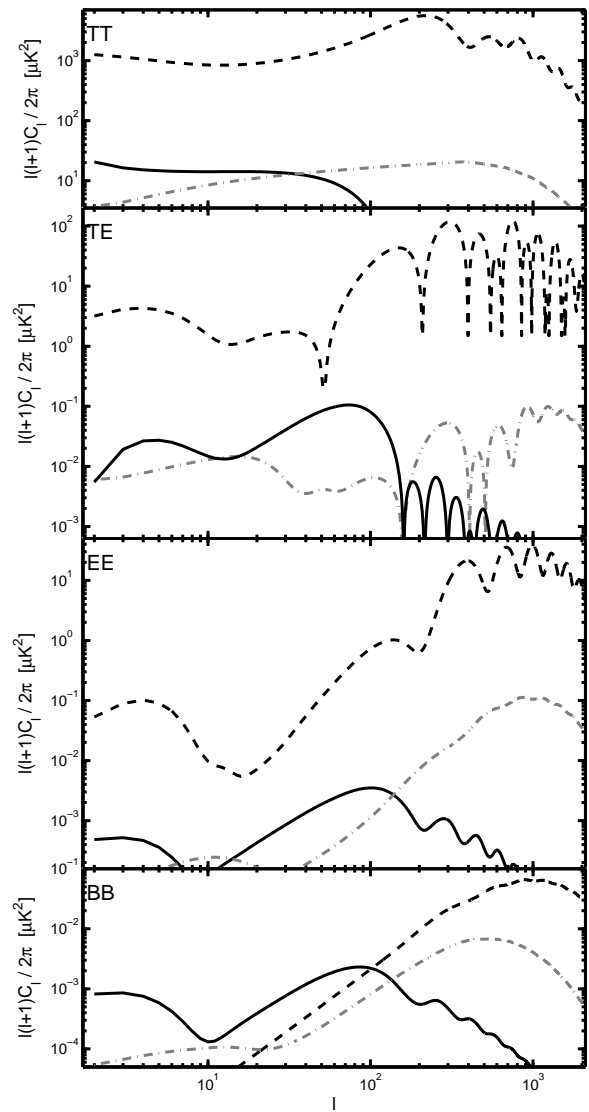}}
\caption{\label{pol}The \CMB temperature and polarization power
spectra contributions from inflationary scalar
modes (black, dashed), inflationary tensor modes (black, solid), and cosmic strings (gray, dot-dashed) 
\cite{Bevis:2007qz}. The inflationary tensors have $r=0.04$ while the string contribution has fractional power at
$\ell=10$ of $\fd=0.01$.}
\end{figure}

\begin{figure*}[t]
\begin{center}
\includegraphics[width=15cm]{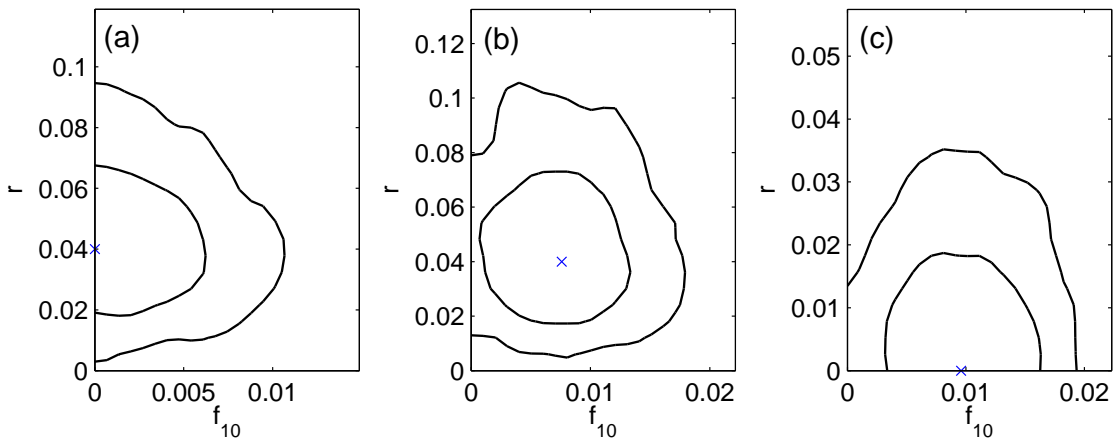}
\caption{\label{fig1} $68\%$ and $95\%$ contours of the marginalized
2D posterior distribution for a cosmological model with (a) $r=0.04$
and no string component,
(b) $r=0.04$ and $\fd=0.008$, 
and (c)
$\fd=0.01$ with no tensor component. The actual models are depicted with a cross.}
\end{center}
\end{figure*} 
 
We show the
contributions to the temperature and polarization power spectra
coming from inflation, strings and tensors in Fig.~\ref{pol},
based on Ref.~\cite{Bevis:2007qz}. The normalizations of
these three components are free parameters, and in this figure are
chosen as follows: the normalization of the inflationary scalar component is
chosen to be the one that matches current \CMB data without including
strings or tensor modes. The string contribution is set at
the $\fd=0.01$ level and the inflationary tensor mode normalization
corresponding to a tensor-to-scalar ratio of $r=0.04$ (at
comoving wavevector $k_{0}=0.01$ Mpc$^{-1}$).\footnote{We define $r$
following the convention of the WMAP papers.} These levels of $\fd$ and $r$ 
are
the typical values we will use in our analysis. 
Tensors and strings are
subdominant in the TT, TE and EE cases (where the data is more
constraining) and it is due to this subdominant nature that one may
wonder whether Planck data will be able to distinguish between
them. By contrast, both tensors and strings dominate in the BB case,
whereas (scalar) inflationary modes only enter through lensing.

We simulate data for a set of different cosmologies, varying the
amount of primordial tensors $r$ and cosmic strings $\fd$.  The values
of $r$ chosen for the fiducial cosmologies lie towards the upper bound
of detection of Planck, rather than the values of $r\sim 10^{-23}$
that string theory seems to suggest. If Planck does detect some extra
ingredient beyond the standard (scalar) concordance model, the
parameter values that would be inferred are at the same level as the
ones considered in this article.

\section{Results and Discussion}

Figure \ref{fig1} shows constraints on tensors and strings when both
these components were fitted for in three different choices of input
cosmology. They show that there is no significant correlation or
degeneracy between the two components; the anticorrelation between $r$
and $\fd$ is just a few percent.  Accordingly, Planck's ability to
measure $r$ is not degraded by allowing the possibility of strings, 
and vice versa. (Here and throughout the other parameters being
varied are the matter and baryon densities, the angular distance
to rescattering, the reionization optical depth, the scalar spectra
index, and the amplitude of primordial density perturbations.)

This exercise showed that trying to fit the fiducial cosmologies with
the correct parameters is very successful, and no degeneracies are
found. However, let us suppose that the actual cosmological model
includes some signal from cosmic strings, but we only try to fit the
data with a model with tensors, or vice versa for the case where the
true model has gravitational waves and no strings.  Will Planck data
be good enough to show that we are trying to fit with the wrong set of
parameters?

\begin{figure}[t]
\begin{center}
\includegraphics[width=8cm]{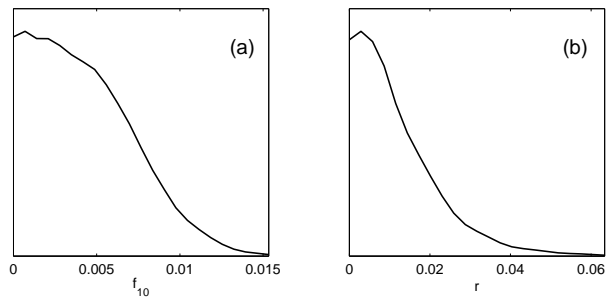}
\caption{\label{fig2} Marginalized 1D distribution (a) for $\fd$
obtained for a fiducial model with $r=0.04$ and no string component
(b) for $r$ obtained when the fiducial model has an $\fd=0.01$ and no
tensors component. The plots shows that fitting for only the wrong
component does not result in any detection but only upper limits.
}
\end{center}
\end{figure}

In order to answer that we created a fiducial model with tensors
$r=0.04$ and no strings, and tried to fit it with a model with no
tensors but strings. In this case (Fig.~\ref{fig2}) no strings are
detected: instead, upper limits are obtained on $\fd$ similar to, but
weaker than, those obtained when we fitted for both components for
this same true model (see Table~\ref{data}). Similarly, the results of
the true model being $r=0$ and $\fd=0.01$, but fitting for just $r$, 
are shown in Fig.~\ref{fig2}. Once again, no $r$ is detected, upper
limits are obtained. We conclude that one detection will not be
mistaken for another.

\begin{table*}[!htb]
\begin{tabular}{|c|c|c|c|c|}
\hline
&  Mean &  Stand. Dev. & $68\%$ Upper Bound& $95\%$ Upper Bound \\
\hline
Fitting for $\fd$ only & 0.0043 & 0.0029 & 0.0056 &  0.0098\\
\hline
Fitting for both $\fd$ and $r$  & 0.0033 & 0.0026 &  0.0041 &
0.0084 \\ 
\hline \hline
Fitting for $r$ only & 0.012 & 0.010 & 0.015 &
0.033\\\hline 
Fitting for both $r$ and $\fd$ & 0.011 & 0.0091 & 0.013 &
0.029\\\hline 
\end{tabular}
\caption{\label{data} First two rows: Values of $\fd$ obtained when
trying to fit a fiducial model with tensors $r=0.04$ and no
strings. Last two rows: Values of $r$ when trying to fit a fiducial
model with strings with $\fd=0.01$ and no tensors.}
\end{table*}

To determine whether most of the string detection capability comes
from Planck's temperature or polarization spectra, we did a similar
analysis using data from only some of the power spectra. 
First we chose not to use BB data. 
The only appreciable  
difference is that the value of $r$ is less constrained.
However, there is still no degeneracy between tensors and strings.

We then also performed the analysis using only temperature data. 
We find that temperature and
polarization offer similar constraining powers. However, with only
temperature there is a positive degeneracy between the scalar spectral
index $n_{\rm s}$ and $r$ and a negative degeneracy between $n_{\rm
s}$ and $\fd$. These degeneracies go away upon adding polarization
data. 
This implies that the current ambiguity that exists between
whether the WMAP data should be interpreted as providing evidence for
$n_{\rm s} \neq 1$ or for strings will not remain when polarization
data improve.

\section{Conclusions}

Our analysis shows that at Planck sensitivity there is no
significant  degeneracy between tensors and cosmic strings. When a set of
cosmological data are fitted using both components, the true input
value of any component is correctly recovered if it is detectable.  If
only one component is fitted and it is the wrong one, then it is not
detected nor misidentified, and upper limits are found similar to, but
weaker than, in the case when both components are fitted. These weaker
bounds are obtained because larger amounts of the wrong component are
required because the other is not being fitted. With actual data, one
would carry out a Bayesian model selection analysis to assess which
was the preferred model to fit, and derive upper limits using Bayesian
model averaging as in Ref.~\cite{LMPW}.

\begin{acknowledgments}

We acknowledge support from PPARC/STFC (N.B., M.H, M.K., A.R.L., P.M.) 
and Marie Curie Intra-European Fellowship
MEIF-CT-2005-009628 (J.U.). This work was partially supported 
by the National Science Foundation under Grant No. PHY05-51164
(M.H.),
Basque Government (IT-357-07), the Spanish Consolider-Ingenio 2010
Programme CPAN (CSD2007-00042) and  FPA2005-04823 
(J.U.).

\end{acknowledgments}

\end{document}